\documentclass[useAMS,usenatbib]{mn2e}

 \usepackage{graphicx}

\title[Deuterium fractionation on
grains]{Deuterium fractionation on interstellar grains studied
with the direct master equation approach}
\author[Tatiana Stantcheva and Eric Herbst]{Tatiana Stantcheva$^{1}$ and Eric
Herbst$^{2}$\thanks{E-mail: herbst@mps.ohio-state.edu (EH)}\\
$^{1}$Department of Physics, The Ohio State University, Columbus,
OH 43210, USA\\
$^{2}$Departments of Physics and Astronomy, The Ohio State
University, Columbus, OH 43210, USA\\}
\begin{document}


\pagerange{\pageref{firstpage}--\pageref{lastpage}} \pubyear{2002}

\maketitle

\label{firstpage}

\begin{abstract}
We have studied deuterium fractionation on interstellar grains
with the use of an exact method known as the direct master
equation approach. We consider conditions pertinent to dense
clouds at late times when the hydrogen is mostly in  molecular
form and a large portion of the gas-phase carbon has already been
converted to carbon monoxide. Hydrogen, oxygen, and deuterium
atoms, as well as CO molecules, are allowed to accrete onto dust
particles and react there to produce various stable molecules. The
surface abundances, as well as the abundance ratios between
deuterated and normal isotopomers, are compared with those
calculated with the Monte Carlo approach. We find that the
agreement between the Monte Carlo and the direct master equation
methods can be made as close as desired. Compared with previous
examples of the use of the direct master equation approach, our
present method is much more efficient. It should now be possible
to run large-scale gas-grain models in which the diffusive dust
chemistry is handled ``exactly''.
\end{abstract}

\begin{keywords}
 molecular processes -- ISM: clouds -- ISM:
molecules.
\end{keywords}

\def\ra{$\longrightarrow$}

\section{Introduction}

Rate equations have long been used in modelling the surface
chemistry that occurs on cold dust particles in the interstellar
medium \citep*{pickles-williams,hasegawa}. It has been pointed out
\citep*{rodgers} that, in the case when the average number of
reactive species on a grain is sufficiently small, the rate
equations may not be suitable to describe the system, since they
do not take into consideration its discrete nature. One
alternative is to use a Monte Carlo approach
\citep{th,rodgers,caselli:deut,charn}.

Recently another approach to this problem \citep{green, biham},
has been proposed. Like the Monte Carlo method, this approach is
based on a master equation.  But, unlike the Monte Carlo
technique, the master equation is directly converted into
differential equations. In particular, the rate equations for the
population of highly reactive species with small surface
abundances are replaced by sets of differential equations for the
probability that certain number of these atoms or molecules exist
on the surface at the same time. One advantage of this new exact
approach is its easy coupling with the differential rate equations
used to model the gas-phase chemistry.  A comparison of the two
approaches is to be found in \citet*{me}.

We have, so far, successfully applied the direct master equation
approach to a system consisting of H and O atoms and CO molecules
accreting onto a grain surface, and reacting to form a variety of
reaction products \citep{me}. In this paper, we report an
expansion of the system to include D atoms accreting onto the
surface and reacting with the other  reactive surface species to
produce deuterated isotopomers. The system is identical to that
studied in \citet{caselli:deut}  with modified rate equations and
a Monte Carlo approach. A somewhat smaller network of reactions
was utilised by \citet{rodgers}.  As opposed to
\citet{caselli:deut}, we concentrate on the doubly, triply and
quadruply-deuterated species which we believe are likely
candidates for observation in highly fractionated low-mass
protostellar sources \citep{parise}.  The main purpose of this
paper; however, is to show how the direct master equation method
can be used efficiently.

The paper is organised as follows. First, in
Section~\ref{overview} a short description of the system under
investigation is given. The theory, including the formulae used in
the calculations, is given in Section~\ref{theory}.
Section~\ref{results} presents the results, and is followed by a
discussion in Section~\ref{discussion}.

\section{The chemical network and physical conditions}
\label{overview}

We have considered a system in which H, O, and D atoms and CO
molecules accrete onto the grain surface and react to form the
stable molecules H$_2$, HD, D$_2$, H$_2$O, HDO, D$_2$O, H$_2$CO,
HDCO, D$_2$CO, CH$_3$OH, CH$_3$OD, CH$_2$DOH, CH$_2$DOD,
CHD$_2$OH, CHD$_2$OD, CD$_3$OH, CD$_3$OD, O$_2$, CO$_2$, and the
highly reactive radicals OH, OD, HCO, DCO, H$_3$CO, H$_2$DCO,
HD$_2$CO, and D$_3$CO. The system pertains to old dense
interstellar clouds, in which hydrogen is mostly molecular and
carbon is mostly in the form of CO.  The calculations were
performed mainly at a temperature T=10 K, for a period of 10$^4$
years, and fixed gas-phase abundances of the accreting species, an
approximation justified by the relatively short period of
evolution of the surface chemistry compared with the time scale of
the gas-phase chemical processes. At times even earlier than
10$^4$ yr,  a steady-state condition is reached in which the
surface populations of the stable species grow linearly with
increasing time while the reactive species have fixed populations.
The number of available sites per grain, which helps to define the
diffusion rates, was taken to be $10^6$; this number refers to
so-called classical grains with a size of 0.1 $\mu$m.

Three different sets of values for the gas-phase densities of H,
O, CO, were used throughout the calculations. Given in Table
~\ref{density}, these sets of values are referred to as the low,
intermediate, and high density cases because they were derived
from gas-phase models at total densities of 10$^3$, 10$^4$, and
10$^5$ cm$^{-3}$, respectively. In all cases, the concentration of
H is near 1 cm$^{-3}$. Unless stated to the contrary, all results
are for a gas-phase abundance for  D of 0.3 cm$^{-3}$. This very
high value relative to the atomic H abundance is presumably
produced via fractionation in the cold gas, although current
gas-phase models cannot easily reproduce such a value
\citep*{roberts}.  Variations in the atomic deuterium abundance
are also considered.

\begin{table}
\caption[]{H, O, and CO gas-phase abundances (cm$^{-3}$)
utilised}\label{density}
\begin{tabular}{clll}
\hline
Abundance $n$ & Low & Intermediate & High \\
\hline
    H & 1.15 & 1.15 & 1.10 \\
    O & 0.09 & 0.75 & 7.0\\
    CO & 0.075 & 0.75 & 7.5\\
    D & 0.3 & 0.3 & 0.3 \\
\hline
\end{tabular}
\end{table}

In calculating the accretion, desorption, and diffusion rates of
the surface species, we followed the methods used in
\citet{hasegawa} -- see also \citet*{caselli:deut, caselli} -- and
used their values for the parameters necessary for the
calculations. All particles were considered to diffuse over the
surface solely via thermal hopping except for H and D atoms, for
which quantum tunnelling was also considered. The rates used for
diffusion are the so-called fast rates \citep{me} because these
magnify the differences between exact and approximate methods for
studying diffusive surface chemistry. Table~\ref{rates} shows the
accretion, evaporation, and diffusion rates over an entire grain
for the accreting species in the model. Note that at the
temperatures considered, the other heavy species in the model do
not diffuse or evaporate at non-negligible rates.

\begin{table}
\caption[]{Assorted rates for selected species at 10
K}\label{rates}
\begin{tabular}{llll}
\hline
                Species  & $k_{\rm acc}$ (cm$^{3}$s$^{-1}$)
        & $t_{\rm evap}^{-1}$ (s$^{-1}$)
        & $t_{\rm diff}^{-1}$ (s$^{-1}$) \\
\hline
H          & 1.45(-5)   & 1.88(-3)  & 5.14(+4) \\
D      & 1.02(-5)   & 1.67(-4)  & 3.92(+2) \\
O          & 3.62(-6)& 2.03(-23) & 4.24(-5)\\
CO         & 2.73(-6) & &\\
\hline
\end{tabular}
\end{table}

The surface reactions used in our model as well as any non-zero
activation energies $E_{\rm a}$ are given in Table
~\ref{reactions}. More details can be found in
\citet{caselli:deut}.

\begin{table}
\caption[]{Surface reactions in the H,O,CO model.}
\label{reactions}
\begin{minipage}{12 cm}
\begin{tabular}{llcll}
\hline Number & \multicolumn{3}{c}{Reaction} & $E_{\mathrm a}$ (K)
\footnote {See \citet{caselli:deut} }    \\
\hline
1  & H + H & \ra & H$_{2}$      &      \\
2  & H + O & \ra & OH       &      \\
3  & H + OH & \ra & H$_{2}$O        &      \\
4  & H + CO & \ra & HCO     & 2000 \\
5  & H + HCO & \ra & H$_{2}$CO  &      \\
6  & H + H$_{2}$CO & \ra & H$_{3}$CO    & 2000 \\
7  & H + H$_{3}$CO & \ra & CH$_{3}$OH   &      \\
8  & H + D & \ra & HD       &      \\
9  & H + OD & \ra & HDO     &      \\
10 & H + DCO & \ra & HDCO       &      \\
11 & H + HDCO & \ra & H$_2$DCO  & 1965  \\
12 & H + D$_2$CO & \ra & HD$_2$CO   & 1925  \\
13 & H + H$_2$DCO & \ra & CH$_2$DOH &   \\
14 & H + HD$_2$CO & \ra & CHD$_2$OH &   \\
15 & H + D$_3$CO & \ra & CD$_3$OH   &   \\
16 & O + O & \ra & O$_{2}$      &      \\
17 & O + CO & \ra & CO$_{2}$    & 1000 \\
18 & O + HCO & \ra & CO$_{2}$ + H   &      \\
19 & O + D & \ra & OD       &   \\
20 & O + DCO & \ra & CO$_2$ + D &   \\
21 & OH + D & \ra & HDO     &   \\
22 & CO + D & \ra & DCO     & 1930  \\
23 & HCO + D & \ra & HDCO       &   \\
24 & H$_2$CO + D & \ra & H$_2$DCO   & 1799  \\
25 & H$_3$CO + D & \ra & CH$_3$OD   &   \\
26 & D + D & \ra & D$_2$        &   \\
27 & D + OD & \ra & D$_2$O      &   \\
28 & D + DCO & \ra & D$_2$CO    &   \\
29 & D + HDCO & \ra & HD$_2$CO  & 1758  \\
30 & D + D$_2$CO & \ra & D$_3$CO    & 1713  \\
31 & D + H$_2$DCO & \ra & CH$_2$DOD &   \\
32 & D + HD$_2$CO & \ra & CHD$_2$OD &   \\
33 & D + D$_3$CO & \ra & CD$_3$OD   &   \\
\hline \\
\end{tabular}
\end{minipage}
\end{table}

\section{Master equation for the system}
\label{theory}

 In the direct master equation method,
 the system corresponding to the grain surface is
represented by a multitude of states that represent different
possible populations of the surface species. In one state we might
have 0 particles of the species A, 1 particle of the species B, 2
particles of the species C, \ldots, whereas a different state
might consist of 1 particle of A, 0 of B, 2 of C, \ldots. With
each state, we associate a probability for the system to be in
this state, and we develop equations for the rate at which these
probabilities change (the master equation).  Basic to the approach
is the fact that the populations of some of the surface species
are correlated. As a consequence, the method, at least formally,
requires the consideration of the evolution of the system as a
whole, as opposed to the rate equation approach, in which the
evolution of the average population of any given species is
followed separately.

In the most general case,  all the surface species are taken into
consideration in each state of the system. Since the standard rate
equations, however, give a sufficiently accurate description of
the evolution of high-abundance species \citep{me}, the inclusion
of these species in the master equation will impose an
unnecessarily heavy load on the computing resources given the need
to integrate many coupled equations simultaneously.  Thus, a more
practical approach involving fewer coupled equations is to include
only the highly reactive and low-abundance surface species in the
master equation, and to use rate-like equations to solve for the
surface populations of the rest of the species. As a guide to
determine which species should be included in the master equation,
one can use the results obtained via rate equations and, if
available, Monte Carlo simulations. More generally, these species
are reactive atoms and radicals, especially those that are
precursors of major grain species. Here we include the 11 species
H, O, D, OH, OD, HCO, DCO, CH$_3$O, CH$_2$DO, CHD$_2$O, and
CD$_3$O. We enumerate these from 1 to 11 and refer to them as
``probabilistic''. The remainder of the species are referred to as
normal and are treated via rate-like equations.

Once we have determined the probabilistic species, we assign a
probability  $P(i_{1},\ldots,i_{11})$ to any state
\{$i_{1},\ldots,i_{11}$\} that consists of $i_{1}$ H atoms,
$i_{2}$ O atoms, $i_{3}$ D atoms, $i_{4}$ OH molecules, \ldots,
$i_{11}$ CD$_3$O molecules, and solve the master equation, which
consists of equations for the rate of change of the state
probabilities $P(i_{1},\ldots,i_{11})$.  These equations are of
the form

\begin{equation}
\label{prob}
\begin{array}{l}
\frac{{\textstyle dP}}{{\textstyle dt}}(i_{1},\ldots,i_{11}) = \\ \\
    \displaystyle{\sum_{\{\mathrm {X}\}}} k_{\mathrm{acc}}
        (\mathrm{X})n(\mathrm{X})
        \left[ P(...,i_{j}-1,...) - P(...,i_{j},...)\right] \\ \\
    + \displaystyle{\sum_{\{\mathrm{X}\}}}
        t_{\mathrm{evap}}^{-1}(\mathrm{X})
        \left[(i_{j}+1)P(...,i_{j}+1,...) -
        i_{j}P(...,i_{j},...)\right] \\ \\
    + \displaystyle{\sum_{\{\mathrm{X,Y}\}}} k_{\mathrm{X},\mathrm{Y}}
    (i_{j}+1)(i_{k}+1)P(...,i_{j}+1,...,i_{k}+1,...) \\ \\
    - \displaystyle{\sum_{\{\mathrm{X,Y}\}}} k_{\mathrm{X},\mathrm{Y}}
    (i_{j})(i_{k})P(...,i_{j},...,i_{k},...) \\ \\
    + \displaystyle{\sum_{\{\mathrm{X}\}}} k_{\mathrm{X},\mathrm{X}}
    \frac{(i_{j}+2)(i_{j}+1)}{2}P(...,i_{j}+2,...) \\ \\
    - \displaystyle{\sum_{\{\mathrm{X}\}}} k_{\mathrm{X},\mathrm{X}}
    \frac{i_{j}(i_{j}-1)}{2}P(...,i_{j},...) \; .
\end{array}
\end{equation}
\noindent

In eqs.~(\ref{prob}), $i_{j}$ corresponds to the surface abundance
of the species X, $i_{k}$ to that of the species Y, $n$(X) is the
gas-phase abundance of X, and $k_{\mathrm{acc}}$(X),
$t_{\mathrm{evap}}^{-1}$(X), and $k_{\mathrm{X},\mathrm{Y}}$ are
the accretion and evaporation rates of X, and the rate coefficient
for reaction between X and Y, respectively \citep{hasegawa}. In
the units used here \citep{caselli}, the rate coefficient is
simply the sum of the the diffusion rates of X and Y over the
entire grain. The first two terms on the right-hand side of
eq.~(\ref{prob}) represent the changes in the probability with
time due to accretion and evaporation processes. The subsequent
terms account for the changes due to reactions between two
probabilistic species, both non-identical and identical. It is
important to note that there should be two more terms on the
right-hand side of eq.~(\ref{prob}), which are present in our
calculations but are not given here for considerations of
simplicity. The first of these terms takes into account the change
of the probability due to reactions between a probabilistic and a
normal species, such as H+CO$\longrightarrow$HCO. The second
accounts for (slow) reactions between two normal species, which
have at least one probabilistic species as a product. In our
current model there are no reactions of the latter type. In
addition to these simplifications, we have also not indicated
changes to more than two stochastic species in any term although
there are reactions in which the populations of three such species
can change (e.g. O + HCO).

The solution of eqs.~(\ref{prob}) gives the probabilities for all
the states, which, in turn, allows the calculation of the average
abundance, or number of species per grain, of each probabilistic species, $\langle
N_{\mathrm{X}}\rangle$, as well as any necessary correlation
terms, of the form $\langle N_{\mathrm{X}}N_{\mathrm{Y}}\rangle$.
These entities, then, are substituted into the equations used to
solve for the abundances of the normal species. Although the
latter equations are very similar to the widely used rate
equations, they bear important differences. Equations~(\ref{reH2})
and~(\ref{reH2CO}) are the expressions for the average abundances
of H$_2$ and H$_2$CO (both being normal species):

\begin{eqnarray}
\label{reH2} \frac{{\textstyle d \langle
   N_{\mathrm{H}_{2}}\rangle}}{{\textstyle dt}} & = &
    - \; t_{\rm evap}^{-1}(\mathrm{H}_{2}) \times \langle N_{\mathrm{H}_{2}}
    \rangle  \nonumber \\
    & & + \; 0.5\; k_{\mathrm{H},\mathrm{H}} \times \langle
    N_{\mathrm{H}}(N_{\mathrm{H}}-1)\rangle,\\
\label{reH2CO} \frac{{\textstyle d \langle
    N_{\mathrm{H}_{2}\mathrm{CO}}\rangle}}{{\textstyle dt}} & = &
    + \; k_{\mathrm{H},\mathrm{HCO}} \times \langle
    N_{\mathrm{H}}N_{\mathrm{HCO}}\rangle \nonumber \\
    & & - \; k_{\mathrm{H},\mathrm{H}_{2}\mathrm{CO}} \times \langle
    N_{\mathrm{H}}\rangle \times \langle N_{\mathrm{H}_{2}\mathrm{CO}}\rangle.
\end{eqnarray}

It can be seen that, unlike regular rate equations, both
correlated abundances and terms in which 1 is subtracted from an
abundance appear.  The rate-like equations are obtained by
summation over single and correlated probabilities for the
relevant stochastic species involved in the formation and
depletion of major species \citep{biham,green,me}.


To propagate the system forward in time, both the master equation
and the rate-like equations must be solved simultaneously.   Since
the probabilistic species have very low abundance, the probability
for the system to be in states that contain high abundances of
these species must be very small. It is reasonable, then, to
neglect such states and limit our consideration to those that
comprise only very few particles. For this purpose, we choose a set of
parameters, \{$N_1$,$N_2$,\ldots, $N_{11}$\} and
$N_{\mathrm{tot}}$, such that we neglect every state,
\{$i_1$,$i_2$,\ldots,$i_{11}$\}, for which the following
conditions hold true:

\begin{eqnarray}
\label{req:1}
 i_j & > & N_j \;\;\;, \mathrm{for}\;\mathrm{any}\;j,\;
\mathrm{or}  \\
\label{req:2}
 \sum_{j} i_j & > & N_{\mathrm{tot}} \;.
\end{eqnarray}

Because the first three particles in
\{$i_1$,$i_2$,\ldots,$i_{11}$\} correspond to H, O, and D, we
require that $N_1$,$N_2$, and $N_3$ be equal to at least 2 so that
molecular hydrogen, oxygen and deuterium can be produced. Thus,
the set of minimum values for the upper limits $N_j$ is
\{2,2,2,1,1,\ldots,1\}. If we choose \{$N_j$\} to be equal to this
set of minimum values,  condition~(\ref{req:1}) will eliminate all
but $\mathcal{N} = 3^3 \times 2^8=6912$ possible states of the
system. This is still a very large number of coupled differential
equations!  A strong reduction in this number, however, can be
obtained by applying condition~(\ref{req:2}), {\em which we have
not done previously}. Of course, the number of states will be
dependent on the value of $N_{\mathrm{tot}}$ chosen. Clearly, any
meaningful value of $N_\mathrm{tot}$ must be larger than or equal
to each $N_j$ and smaller than their sum, $\sum_j{N}_j$. If
$N_\mathrm{tot}$ is taken to be 2, the number of states will be 70
for the set of limits \{2,2,2,1,1,\ldots,1\}.

To understand how this number is determined, consider the
following argument.  Let the number of probabilistic particles be
$m$. For the chosen $N_j$ and $N_{\mathrm{tot}}$, there will be
four different groups of states: the first group, [0,0,\ldots,0],
consisting of only one state \{0,0,\ldots,0\}, the second group,
[1,0,\ldots,0], consisting of $m$ different states, the third
group, [1,1,0,0,\ldots,0], consisting of $\frac{m(m-1)}{2!}$
states, and the fourth group, [2,0,\ldots,0], consisting of only 3
states -- \{2,0,\ldots,0\}, \{0,2,0,\ldots,0\}, and
\{0,0,2,0,\ldots,0\}. In our model $m$=11, so that the total
number of states is $1+11+55+3=70$.

Although  analytical formulae can be derived for the number of
states determined by other specific values for $m$,
\{$N_1$,$N_2$,\ldots,$N_{\mathrm m}$\} and $N_{\mathrm{tot}}$, we
choose to determine this number by a computer subroutine.

\section{Results}
\label{results}

 Table~\ref{abund} shows a comparison between the surface
 populations (in monolayers per grain)
of  normal (stable, non-deuterated) species calculated via the
direct master equation (ME) and the Monte Carlo (MC) approaches
for the low, intermediate, and high density cases. To use the
former method, all $N_j$ were set to their minimum values except
for $N_2$ and $N_4$, which correspond to O and OH. The
calculations showed the latter two average abundances to be close to unity
in some cases so that higher values for their upper limits had to
be considered.

\begin{table*}
 \caption[]{Populations in mono-layers at 10$^{4}$ yr and 10 K.}\label{abund}
\begin{tabular}{l|rr|rr|rr}
\hline Species & \multicolumn{2}{c|}{High Density} &
\multicolumn{2}{c|}{Interm. Density} & \multicolumn{2}{c}{Low Density} \\
\hline
    & MC & ME & MC & ME & MC & ME\\
\hline
CO  & 5.0 & 4.9 & 0.00 & 0.00 & 0.0 & 0.0 \\
H$_2$O & 1.4 & 1.2 & 0.45 & 0.45 & 0.070 &
0.069 \\
O$_2$   & 2.7 & 2.4 & 0.080 & 0.079 & 0.0013 &
0.0013\\
CO$_2$  & 0.67 & 0.71 & 0.055 & 0.055 & 7.7(-4)
& 7.2(-4)\\
H$_2$CO & 0.44 & 0.44 & 0.0 & 0.0 & 0.0 & 0.0 \\
CH$_3$OH & 0.088 & 0.095 & 0.42 & 0.42 &
0.046 & 0.046 \\
Total abundance & 11.1 & 10.5 & 1.37 & 1.36 & 0.165 & 0.164
\\
\hline
$N_2$=$N_\mathrm{O}$   & & 4 & & 4 & & 2\\
$N_4$=$N_\mathrm{OH}$   & & 4 & & 4 & & 1 \\
$N_\mathrm{tot}$ & & 4 & & 4 & & 3 \\
$\mathcal{N}$ & & 816 & & 816 & & 265 \\
CPU sec (Cray SV1) & 789 & 313 & 366 & 245 & 366 & 14 \\
\hline
\end{tabular}
\end{table*}

Table~\ref{abund} also shows the values of $N_2$ and $N_4$ for the
three cases as well as the limit $N_{\rm tot}$ for the total
number of probabilistic species and the total number of states
$\mathcal{N}$.  One can see that for the high and intermediate
density cases, the direct master equation method requires the
solution of more than 816 coupled differential equations since the
number $\mathcal{N}=816$ does not include the rate-like
differential equations for the stable species.

For the low and intermediate density cases, the agreement between
the master equation and the Monte Carlo approaches is excellent,
and the master equation approach outperforms the Monte Carlo
calculation in terms of computer time, dramatically so for the low
density limit. For the high density case, the agreement is not as
excellent, but the results of the two methods deviate by at most
10\%. This small deviation can be explained by the high abundance
of O and OH on the surface, a condition that requires that higher
values for $N_2$ and $N_4$, and consequently many more states, be
considered in the master equation approach.  In general, the
agreement between the two formally exact approaches can be made as
good as desired by considering sufficiently high values for the
$N_j$ and for $N_\mathrm{tot}$. This, however, will mean longer
times for the master equation calculations to run, so one needs to
find the right balance between the desired accuracy and the usage
of computing resources.  In general, it is a current weakness of
the master equation approach that there is no obvious algorithm to
determine the proper upper limit to the number of states to be
considered.

The fractionation ratios, {\it f}XD, are defined as the ratio
between the abundances of the deuterated and the normal
isotopomers. The results for the fractionation ratios are given in
Table~\ref{fract} for the three cases discussed previously.  The
agreement between the two methods is reasonable but not perfect;
here the major problem lies in the Monte Carlo approach, which is
not accurate for small populations.  Indeed, for some of the
smaller fractionation ratios, no values could be calculated with
this method since populations less than unity are not allowed. As
an example, consider the case of CHD$_2$OH at intermediate
density. From the tables, one sees that the overall population of
this molecule in the Monte Carlo approach is approximately 140
molecules.  While the random (square-root) error is thus at the 10\% level, the
actual error is considerably greater and is probably large enough
to cover the 30-40\% difference between the two methods of
calculation.

\begin{table*}
\caption[]{Abundance ratios {\it f} at 10$^{4}$ yr and 10 K.}\label{fract}
\begin{tabular}{l|rr|rr|rr}
\hline Species & \multicolumn{2}{c|}{High Density} &
\multicolumn{2}{c|}{Interm. Density} & \multicolumn{2}{c}{Low Density} \\
\hline
    & MC & ME & MC & ME & MC & ME\\
\hline
{\it f}HDCO & 0.33 & 0.34 & \ldots & \ldots & \ldots & \ldots \\
{\it f}D$_2$CO & 0.026 & 0.027 & \ldots & \ldots & \ldots & \ldots \\
{\it f}CH$_3$OD & 0.20 & 0.20 & 0.19 & 0.18 & 0.18 & 0.18 \\
{\it f}CH$_2$DOH & 0.70 & 0.71 & 0.19 & 0.19 & 0.19 & 0.19 \\
{\it f}CH$_2$DOD & 0.14 & 0.14 & 0.034 & 0.034 & 0.034 & 0.034 \\
{\it f}CHD$_2$OH & 0.17 & 0.18 & 3.3(-4) & 2.1(-4) & 2.6(-4) & 1.4(-4) \\
{\it f}CHD$_2$OD & 0.034 & 0.035 & 6.4(-5) & 3.9(-5) & \ldots &
2.5(-5)
\\
{\it f}CHD$_3$OH & 0.014 & 0.015 & \ldots & 5.6(-8) & \ldots &
2.3(-8)
\\
{\it f}CHD$_3$OD & 0.0025 & 0.0029 & \ldots & 1.0(-8) & \ldots &
4.3(-9) \\
{\it f}HDO & 0.39 & 0.38 & 0.38 & 0.38 & 0.39 & 0.39 \\
{\it f}D$_2$O & 0.038 & 0.036 & 0.036 & 0.036 & 0.036 & 0.038 \\
\hline
\end{tabular}
\end{table*}

\begin{figure*}
\vspace{6pc} \rotatebox{-90}{\includegraphics[width=8
cm]{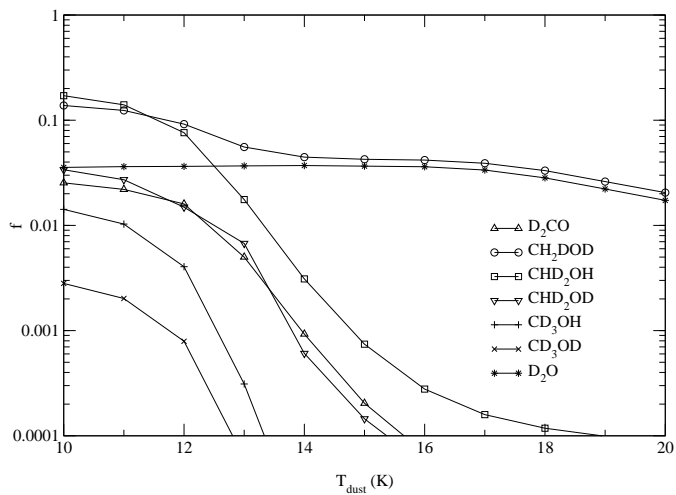}} \caption[]{Fractionation ratios $f$ vs. T$_{\rm
dust}$(K) at 10$^4$ yr for the high density case, with n(D)=0.3
cm$^{-3}$,
 computed with the direct master equation technique. The results
lie close to a simple analytical limit near 10 K. } \label{fvst}
\end{figure*}

\begin{figure*}
\vspace{6pc} \rotatebox{-90}{\includegraphics[width=8
cm]{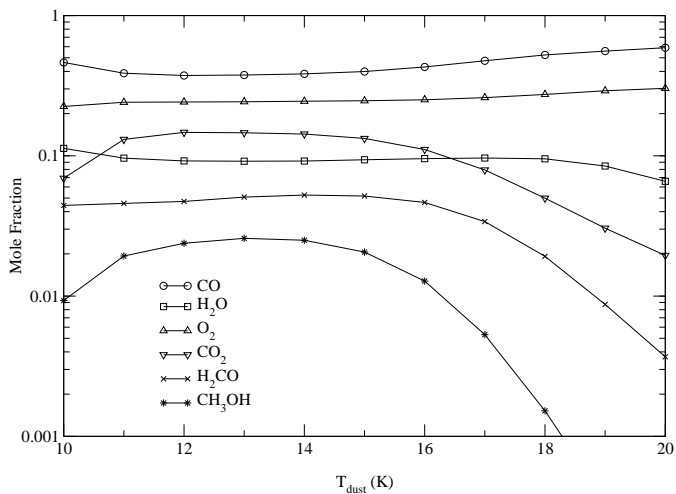}} \caption[]{Mole fractions of major surface species
vs. T$_{\rm dust}$(K) at 10$^4$ yr for the high density case,
computed with the direct master equation technique. } \label{xvst}
\end{figure*}

We have also used the direct master equation approach to model the
chemistry in the temperature range 10-20 K for the high density
case. Fig. ~\ref{fvst} shows the fractionation ratio $f$ of doubly
and multiply deuterated isotopomers in this temperature range,
while a plot of the mole fraction of the normal isotopomers vs.
temperature is given in Fig. ~\ref{xvst}.  The parameters for the
calculation (number of states, etc.) are the same as shown in
Table~\ref{abund}. It can be seen that as the surface temperature
increases from 10 to 20 K, the species (H$_2$CO, CH$_3$OH) made by
hydrogenation involving reactions with activation energy decline
sharply in abundance.  This non-intuitive result derives from the
fact that at higher temperatures, H atoms evaporate before they
react.   In addition, for most species, the fractionation ratios
decline sharply with increasing temperature, especially for
isotopomers that must be formed by reactions with activation
energy and involving D atoms, since these atoms tunnel more poorly
than their H counterparts and are more likely to evaporate.

\begin{figure*}
\vspace{6pc} \rotatebox{-90}{\includegraphics[width=8
cm]{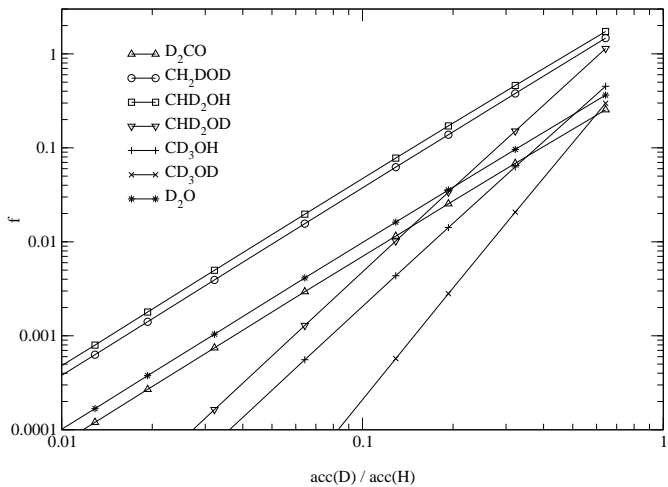}} \caption[]{Fractionation ratios $f$ for high
density conditions at 10 K vs. acc(D)/acc(H, the ratio of the
accretion rate of atomic D to atomic H.} \label{fvsd}
\end{figure*}

Finally, the dependence of the fractionation ratios for doubly and
multiply deuterated isotopomers on the ratio of atomic deuterium
to hydrogen is shown in Fig.~\ref{fvsd} for a 10 K cloud at high
density. The abscissa is actually the ratio of the accretion rate
of atomic D to that of atomic H; this is equal to the ratio of the
gas-phase abundances multiplied by a factor of $2^{-0.5}$ that
derives from the relative speeds of D and H.  Not surprisingly,
the fractionation ratios increase as the D/H accretion ratio
increases, with the slope proportional to the number of deuterium
atoms in the isotopomer.  In fact, the results in Fig.~\ref{fvsd}
lie very close to the very simple limit \citep{tiel, millar} in
which all H and D atoms that accrete onto the surface of a grain
eventually react with the CO reservoir despite the activation
energy barriers.  In this instance, if we let $R$ be the ratio of
the accretion rate of D to that of H, the fractionation ratios for
the deuterated species are simple powers of $R$ multiplied by a
statistical factor, where the power is just the number of D atoms
in the isotopomer.  The statistical factor expresses the number of
possible paths leading to the isotopomer, which, for example, is 3
for CH$_{2}$DOH and 1 for CH$_{3}$OD. This simple limit is
independent of temperature, and so works progressively more poorly
as the temperature is raised from 10 K, as can be seen from the
temperature dependence in Fig.~\ref{fvst}. It also works more
poorly for low and intermediate densities, where the CO reservoir
does not exist.

\section{Discussion}
\label{discussion}

We have shown how a moderately complex system of surface chemical
reactions can be solved by the formally exact direct master
equation method.  This method, or the equivalent Monte Carlo
approach, is needed in the so-called accretion limit, which
pertains to a situation in which the average abundance of
important reactive species on a grain surface is less than unity.
In the accretion limit, normal rate equations are likely to be
inaccurate, and a more exact method, which treats the discrete
nature of the surface populations, is needed.  Of the two exact
methods used to date, the direct master equation approach offers
the possibility of relative ease of implementation in combined
gas-grain approaches since it involves the solutions of coupled
differential equations.

 In order to reach the level of complexity
discussed here, we have introduced a new method to limit the
number of states considered. This goal is achieved by limiting the
{\em total} number of reactive species considered on a grain. In
the current system, which involves 11 reactive species, we have
limited consideration to a total number of species in the range
3-4. Of course, this approach only works successfully if the
probability for even this number of total species is small, which
is what defines the accretion limit.

 If one
eliminates D and the five deuterium-containing radicals from the
11 reactive species considered here, and replaces them with other
reactive species leading to major grain constituents, it is
entirely possible that the master equation treatment described
here can be extended to study combined gas-grain models of
considerable complexity, in which gas-phase abundances change with
time. Problems to be solved before this goal becomes a reality
include the need to develop a criterion for the smallest number of
states that should be considered, and the related need to
determine when and for which species normal rate equations can be
utilized. These problems are certain to depend on physical
conditions such as grain size and temperature, as well as the
rates of diffusion chosen.  It would be ideal if a program could
be developed to decide these issues as the integration with time
proceeds.

Of the deuterated species discussed here, HDCO, D$_{2}$CO,
CH$_{3}$OD, CH$_{2}$DOH, CHD$_{2}$OH, and HDO have already been
detected in interstellar sources.  A low-mass protostellar source
-- IRAS16293-2422 -- contains both deuterated isotopomers of
formaldehyde, both singly deuterated isotopomers of methanol, and
CHD$_{2}$OH in the gas phase \citep{parise}. It is likely that at
least some of the fractionation occurs on interstellar grains and
is followed by evaporation back into the gas \citep{rodgers}. As
shown by \citet{parise}; however, the granular deuteration model
presented here and in \citet{caselli:deut}, as well as the simpler
one in \citet{rodgers}, does not account quantitatively for all of
the observations with any single value of gas-phase D chosen.
Clearly, a more complex gas-grain model is needed, in which
time-dependent fractionation occurs both in the gas and on
dust-particle surfaces.  Our current success with the direct
master equation method may well allow us to consider such a
complex gas-grain system in the future.

\section*{Acknowledgments}
 We thank the National Science Foundation
(US) for support of the Ohio State program in astrochemistry and
the Ohio Supercomputer Center for time on their Cray SV1 machine.
Special thanks go to V. I. Shematovich for carefully reading a
version of the manuscript.

\bsp

\label{lastpage}


\begin{thebibliography}{99}

\bibitem[\protect\citeauthoryear{Biham et~al.}{2001}]{biham}
Biham O., Furman I., Pirronello V., Vidali G., 2001,  ApJ, 553,
595

\bibitem[\protect\citeauthoryear{Brown \& Millar}{1989}]{millar}
Brown P.D., Millar T.J., 2002, MNRAS, 240, 25P


\bibitem[\protect\citeauthoryear{Caselli, Hasegawa \& Herbst}{Caselli et~al.}{1998}]{caselli}
Caselli P., Hasegawa T.I., Herbst E., 1998,  ApJ,495, 309

\bibitem[\protect\citeauthoryear{Caselli et~al.}{2002}]{caselli:deut}
Caselli P., Stantcheva T., Shalabiea O., Shematovich, V.I., Herbst
E., 2002, Plan. Sp. Sci., in press

\bibitem[\protect\citeauthoryear{Charnley}{2001}]{charn}
Charnley S.B., 2001, ApJ, 562, L99

\bibitem[\protect\citeauthoryear{Charnley, Tielens \& Rodgers}{Charnley et al.}{1997}]{rodgers}
Charnley S.B., Tielens A.G.G.M., Rodgers S.D., 1997, ApJ, 482,
L203


\bibitem[\protect\citeauthoryear{Green et~al.}{2001}]{green}
Green N.J.B., Toniazzo T., Pilling M.J., Ruffle D.P., Bell N.,
  Hartquist, T.W., 2001, A\&A, 375, 1111

\bibitem[\protect\citeauthoryear{Hasegawa, Herbst \& Leung}{Hasegawa et~al.}{1992}]{hasegawa}
Hasegawa T.I., Herbst E., Leung C.M., 1992,  ApJS, 82, 167


\bibitem[\protect\citeauthoryear{Parise et al.}{2002}]{parise}
Parise, B. et al., 2002, A\&A, in press

\bibitem[\protect\citeauthoryear{Pickles \& Williams}{1977}]{pickles-williams}
Pickles J.B., Williams D.A., 1977,  Ap\&SS, 52, 443

\bibitem[\protect\citeauthoryear{Roberts, Herbst \& Millar}{Roberts et al.}{2002}]{roberts}
Roberts H., Herbst E., Millar T.J., 2002, MNRAS, 336, 283


\bibitem[\protect\citeauthoryear{Stantcheva, Shematovich \& Herbst}{Stantcheva et~al.}{2002}]{me}
Stantcheva, T., Shematovich V.I., Herbst, E., 2002,  A\&A, 391,
1069

\bibitem[\protect\citeauthoryear{Tielens}{1983}]{tiel}
Tielens A.G.G.M., 1983, A\&A, 119, 177

\bibitem[\protect\citeauthoryear{Tielens \& Hagen}{1982}]{th}
Tielens A.G.G.M., Hagen W., 1982,  A\&A, 114, 245




\end{thebibliography}
\end{document}